# CaregiverConnect: A Mobile Application for Stress Tracking and Support in Dementia Care

Yagna Manasa BOYAPATI[a,1] and Dr. Arshia KHAN[b]

[a,b] *Department of Computer Science, University of Minnesota Duluth, Duluth, MN, USA*

ORCiD ID: Yagna Manasa Boyapati https://orcid.org/0009-0004-1427-3787, Dr. Arshia Khan https://orcid.org/0000-0001-8779-9617

**Abstract.** The rising global prevalence of dementia places escalating demands on paid professional caregivers, who frequently experience severe stress and professional isolation. Existing mobile health (mHealth) applications rarely integrate real-time stress monitoring with multimodal relief strategies. CaregiverConnect, a Flutter/Firebase mobile application comprising nine feature modules, was evaluated in a single-group, pre-post pilot study with 10 paid professional dementia caregivers at two nursing home facilities over two weeks. Descriptive analysis yielded an exploratory effect estimate for self-reported stress change (Cohen's $d = 1.60$; interpreted with caution given $N = 10$ and no control group), an improvement in emotional support responses from 60% to 100% of participants, and near-perfect usability ratings (UI Navigation: 4.90/5.0). These preliminary findings support the feasibility and acceptability of CaregiverConnect and identify end-of-shift stress peaks as a priority target for future adaptive design. Larger randomized controlled trials are needed to establish efficacy.

**Keywords.** Dementia care, mobile health, caregiver support, stress tracking, mHealth, Flutter, Firebase

## 1. Introduction

Dementia affects over 55 million people worldwide, with nearly 10 million new cases annually [1]. Over 21% of Americans serve as caregivers, 70% reporting work-related disruptions, 20% having quit employment, and over 40% reducing to part-time work due to caregiving demands [2]. Approximately 68% of dementia caregivers report high burden and 65% exhibit depressive symptoms, with burden correlating with caregiver sex, income, and education level [3]; up to 75.7% of caregivers report at least one unmet psychological support need [8]. Despite the scale of this workforce challenge, existing support mechanisms remain fragmented and reactive.

Most mHealth applications focus on patient cognition rather than caregiver wellbeing [4], and fewer than 18% address multiple caregiver support domains [5]. Technology-based caregiver interventions have yielded only modest effects on burden (SMD: −0.13) and depression (SMD: −0.20) in meta-analyses [6], underscoring the need for integrated, multimodal solutions. This paper presents CaregiverConnect, a Flutter-based mobile application combining real-time stress tracking, evidence-based relief mechanisms, patient health logging, expert consultation, and peer support, evaluated in a two-week exploratory pilot study with 10 paid professional dementia caregivers [11].

[1] Corresponding Author: Dr. Arshia Khan, akhan@d.umn.edu.

## 2. Methods

*2.1 Application Design and Development and Database Architecture*

CaregiverConnect was built using Flutter SDK (v3.24.3) and Dart (v3.5.3) for cross-platform iOS/Android deployment, with Firebase providing real-time Firestore NoSQL storage, role-based Authentication, and Cloud Functions. The database uses four collections: caregivers (profile, credentials, shift data), patients (medical history, preferences, behavioral notes), stress_logs (caregiver ID, timestamp, time-of-day flag, eight survey responses), and health_logs (fluid intake, toileting, sleep, behavioral observations). Firebase Security Rules restrict each caregiver's read/write access to their assigned patients' records only; all data are encrypted in transit (TLS) and at rest (AES-256). UI was prototyped in Figma; Agile sprint cycles with weekly UAT sessions, where 2–3 users rated each screen on a 1–5 usability scale, with screens below 4.0 revised before the next sprint, guided iterative development [7].

Nine feature modules were implemented: (1) Caregiver Directory; (2) Patient Directory with role-based privacy controls; (3) Expert Help Station linking caregivers with geriatricians, psychiatrists, and social workers; (4) Stress Tracking (Section 2.3); (5) Timesheet and Task Management; (6) Stress Relief including Box Breathing, 4-7-8 Breathing, hydration reminders, mood-based nutrition, cognitive and relaxation games, and music therapy; (7) Family Communication Platform; (8) Patient Health Log; and (9) Resource Library.

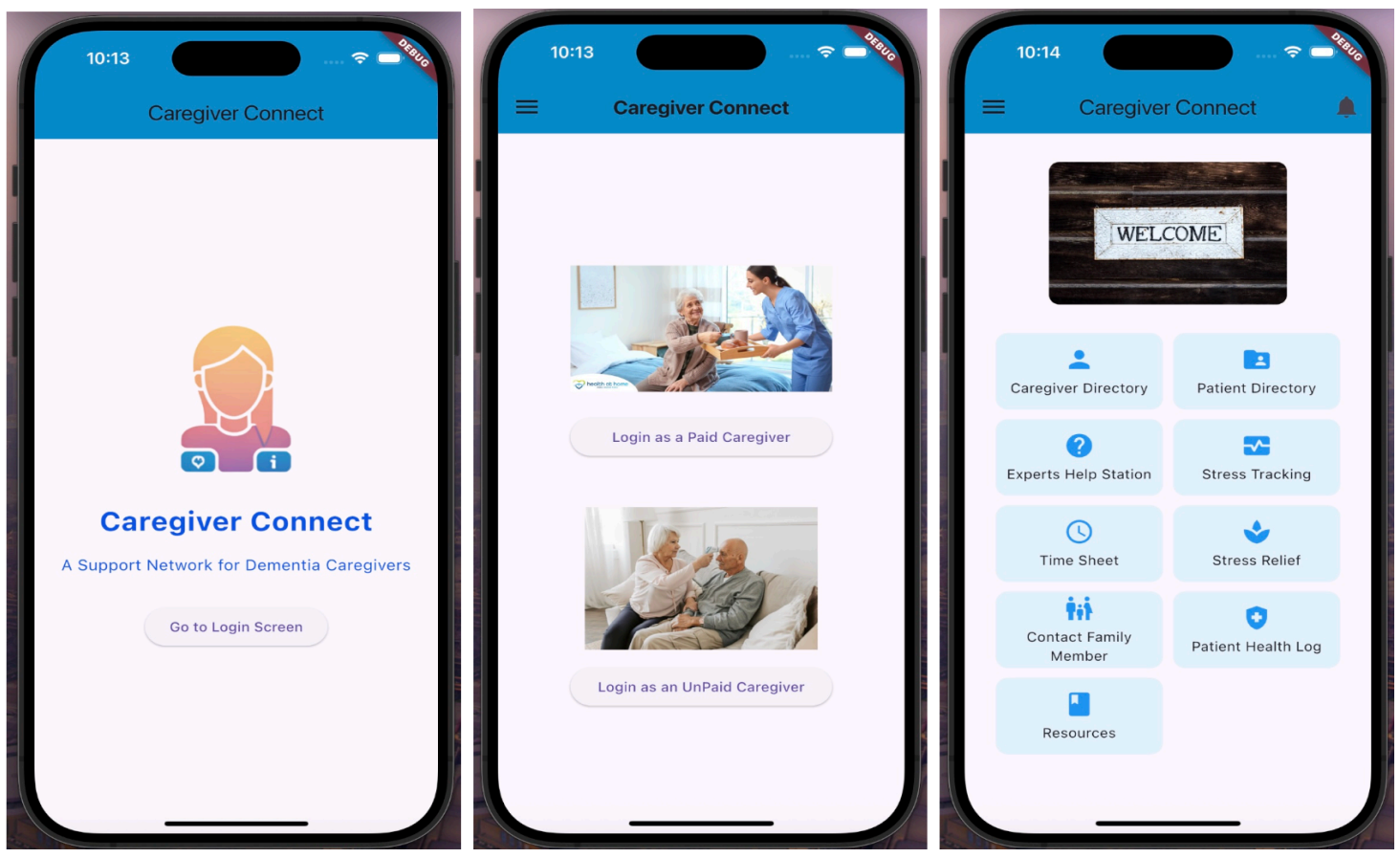

**Figure 1.** Main/Splash Screen, Login and Home Screens of the Caregiver Connect app

*2.2 Study Design and Participants*

A single-group, pre-post pilot design was employed over two weeks at two Minnesota nursing homes: Adult Day Services and Hope House. The target group was defined as paid professional caregivers: certified nursing assistants, licensed practical nurses, or registered nurses, actively employed in dementia care with at least 20 weekly hours of direct patient contact and basic smartphone literacy. Per-diem caregivers were excluded as irregular schedules prevent consistent three-daily survey completion. Ten eligible staff members provided written informed consent following institutional approval at each facility; these same ten individuals formed the study cohort and completed all study procedures. Participants received standardized Google Pixel devices pre-loaded with CaregiverConnect and attended a structured in-person training session covering all nine modules, with particular focus on Stress Tracking and Stress Relief. At study end, devices were returned and the post-study questionnaire completed.

### *2.3 Measurement Instruments and Analysis*

Four instruments were used: (a) *pre-study questionnaire* — baseline stress (1–5 Likert; 1 = no stress, 5 = extreme stress), emotional support (yes / sometimes / no), loneliness frequency, and open-ended questions on challenges. (b) *Eight-item daily stress tracking survey* — a custom, non-validated instrument administered three times daily (morning, afternoon, evening) via embedded Qualtrics, rated on a 1–7 fully labeled scale (1 = not at all, 4 = moderately, 7 = extremely), covering sleep restfulness, energy, physical discomfort, stress, focus, workload manageability, support need, and emotional calmness; the three-daily EMA schedule was adopted from Fuller-Tyszkiewicz et al. [9]. (c) *Post-study questionnaire* — perceived stress reduction, feature usefulness, and scheduling satisfaction (all 1–5; 1 = not at all, 5 = completely); navigation ease (1 = very difficult, 5 = very easy); plus, binary yes / sometimes / no items for expert connection, family communication, and emotional support. (d) *Open-ended responses* analyzed by thematic coding. Only descriptive statistics and an exploratory Cohen's d (pooled SD) were computed; no inferential tests were conducted. All estimates are preliminary given N = 10 and the absence of a control group.

## 3. Results

### *3.1 Participants*

Ten paid professional dementia caregivers completed the full two-week pilot (8 female, 2 male). There was no attrition.

### *3.2 Stress reduction Outcomes*

On the pre-study 1–5 stress scale (1 = no stress, 5 = extreme stress), mean baseline stress was 1.80 (±1.14). On the post-study 1–5 stress reduction scale (1 = no reduction, 5 = complete relief), mean rating was 3.40 (±0.84). Note that these two scales measure different constructs (level vs. perceived reduction) and are not directly comparable; the exploratory Cohen’s d of 1.60 was computed on these paired values as a directional indicator only and is very likely inflated by the small sample and absence of a control condition. The direction of change was consistent across participants: post-intervention distributions showed greater central tendency and reduced variability (Fig. 2).

### *3.3 Feature Effectiveness*

All four features rated above 4.0 on the 1–5 post-study scale: Care Plans/Scheduling (4.80 ± 0.42), UI Navigation (4.90 ± 0.32), Patient Tracking (4.70 ± 0.48), Stress Relief (4.20 ± 0.92). Post-study domain scores—stress relief usefulness, health tracking usefulness, scheduling satisfaction, and navigation ease—all received mean ratings of 5.0/5.0; self-reported stress reduction following app use was rated 4.0/5.0.

### *3.4 Communication and Emotional Support*

Binary post-study items (response options: yes / sometimes / no) showed that all 10 participants responded “yes” to successfully connecting with experts, coordinating with patient families, and feeling emotionally supported and less isolated. Pre-study, 60% of participants had responded “yes” and 40% “sometimes” to the emotional support item;

post-study, 100% responded “yes.” All 10 participants stated they would recommend CaregiverConnect to other caregivers.

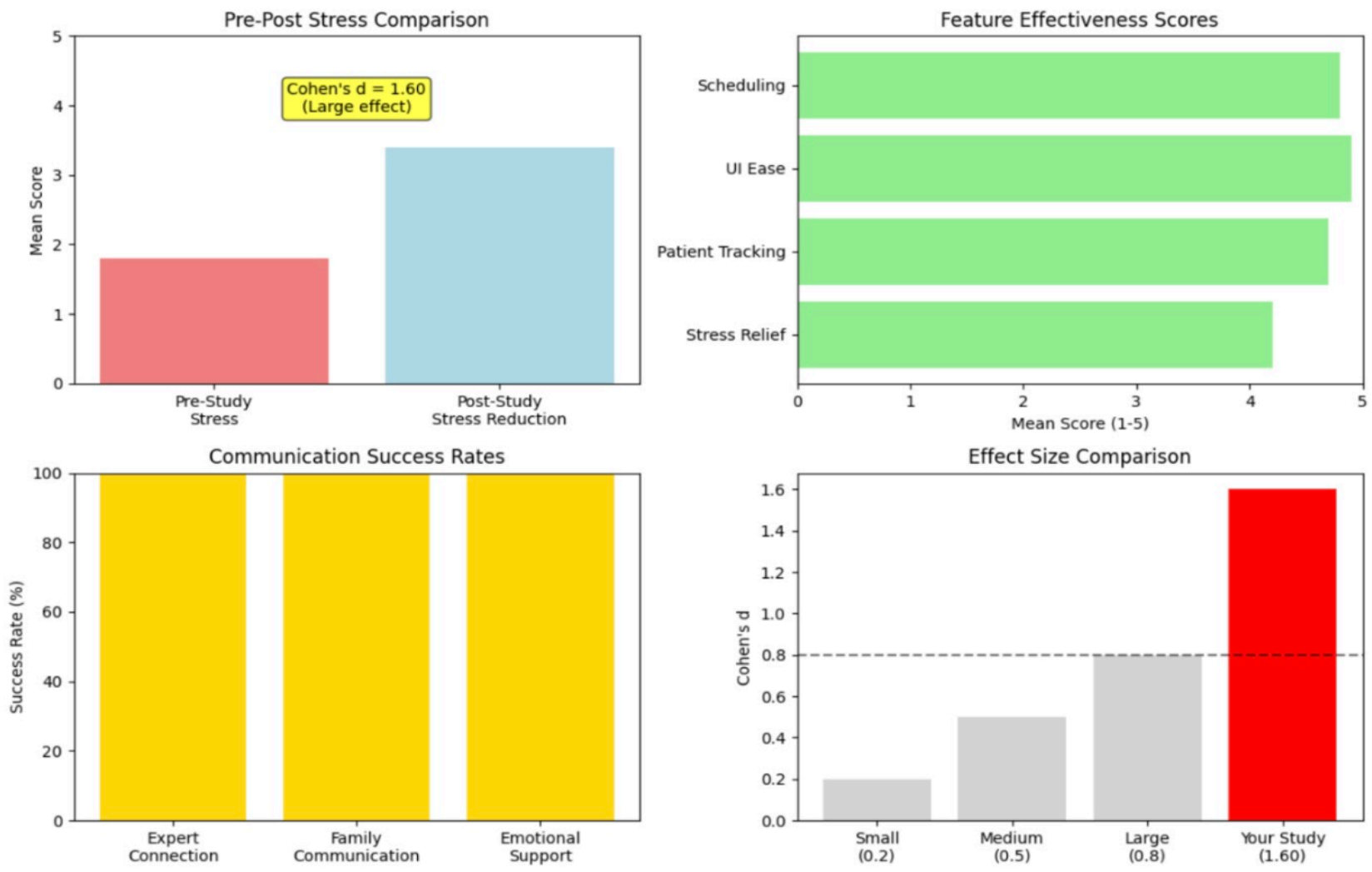

**Figure 2.** Fig. summarizes four outcome panels: pre-post stress scores, feature effectiveness ratings (both 1–5), and exploratory effect size against benchmarks.

*3.5 Temporal Stress Patterns*

On the 1–7 EMA scale, descriptive means showed energy declining from morning (6.2) to evening (5.2) and focus dropping more steeply from morning/afternoon (6.0) to evening (4.2), consistent with accumulating end-of-shift fatigue. Need for support showed the most marked rise across the day, from morning (0.4) to evening (2.0); stress, while low in absolute terms, peaked in the evening (1.8). Emotional calmness remained stable at approximately 6.25 across all time points. These are descriptive observations from a small uncontrolled pilot; no inferential statistics or formal correlation coefficients were computed.

*3.6 Qualitative Findings*

Thematic coding of open-ended pre-study responses identified three challenge themes: time and workload management (reported by 4 of 10 participants), system and coordination problems (3 of 10), and team or resource constraints (2 of 10). In open-ended post-study responses to “which features did you find most helpful and why,” music and entertainment features were cited by 7 of 10 participants; patient health logging and care coordination tools were cited by 3 of 10. Caregivers most frequently attributed their willingness to recommend the application to its low time burden and its role in promoting stress self-awareness.

## 4. Discussion

Preliminary findings indicate CaregiverConnect is feasible and acceptable. The exploratory Cohen's d of 1.60 exceeds SMDs in prior meta-analyses [6] and parallels gains in controlled multimodal trials [10], but with N = 10, no control group, and no inferential testing it almost certainly overestimates the true effect and is not confirmatory. End-of-shift peaks in support need and fatigue point to proactively timed interventions as a key design direction; near-perfect usability [7] and the 60%→100% emotional support shift [8] confirm acceptability. Limitations include the small convenience sample, single-arm design, two-week window, and custom non-validated instrument. Future work should use a validated scale include a randomized control arm, and conduct longitudinal follow-up.

## 5. Conclusion

This pilot study demonstrates the feasibility and acceptability of CaregiverConnect as a multimodal mobile support platform for paid professional dementia caregivers. Preliminary findings show consistent directional improvements in self-reported stress, near-perfect usability scores, and unanimous recommendation intent which provide sufficient grounds to justify a larger, controlled evaluation. Temporal EMA data identify end-of-shift as a high-priority window for adaptive digital intervention.